\documentclass{article}
\usepackage{amsmath}
\usepackage{graphicx}
\begin{document}

\title{Method of images applied to an opto-mechanical Fabry-Perot resonator}
\date{March 16, 2013}
\author{Raymond Chiao (rchiao@ucmerced.edu)}
\maketitle

We show here that effectively relativistic motions of the images of objects in a Fabry-Perot resonator can occur. Since these
images can move relativistically, they can emit radiation that can be seen
by a stationary observer (e.g., the eye in Figure 1).

\begin{figure}[ptb]
\begin{center}
\includegraphics[width=5in]{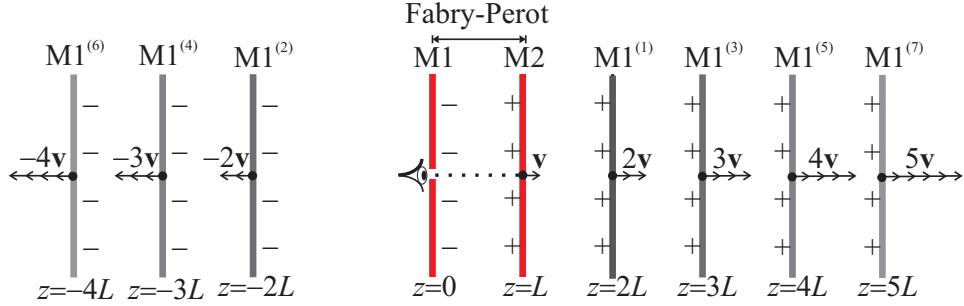}
\end{center}
\caption{Fabry-Perot resonator with one
stationary mirror M1 and one moving mirror M2. The observer (the eye) will see multiple images M1$^{(m)}$, $(m=1,2,3...)$, of the mirror M1 with
steadily increasing velocities as the order $m$ of the image of M1 increases.}
\end{figure}

Consider the opto-mechanical Fabry-Perot resonator operating at microwave
frequencies sketched in Figure 1. Let the axis of the resonator be the $z$
axis, and let the mass of stationary mirror M1 be a large mass $M$, but let
the mass of the moving mirror M2 be a small, mesoscopic mass $m$. For
example, M2 could be a Planck-scale mass of a thin, flexible superconducting membrane
mirror that can easily be driven into motion at microwave frequencies \cite{FQMT11}%
\cite{Aharonov MS}.%

The left mirror M1 of this resonator, being heavy,\ is essentially
motionless, but its right mirror M2, being light, can be moving to the right
relative to the mirror M1 with an instantaneous velocity $\mathbf{v}$, as
seen by the eye in Figure 1. We shall assume that the left mirror M1 will be
negatively charged with a charge $-q$ and that the right mirror M2\ will be
positively charged with a charge $+q$, like in a capacitor. 

The light mirror M2 can be mechanically driven, for example, by some
microwave-frequency Coulomb force $F_{z}\left( t\right) $ due to an
instantaneous longitudinal electric field $E_{z}\left( t\right) $\ of a
transverse magnetic mode of the resonator, via the relationship \cite{FQMT11}%
\cite{Aharonov MS}%
\begin{equation}
F_{z}\left( t\right) =qE_{z}\left( t\right) 
\end{equation}%
The two mirrors M1 and M2 could be composed of superconductors, so that the
quality factor $Q$ of the Fabry-Perot resonator could be extremely high, for
example, on the order of 
\begin{equation}
Q\sim 10^{10}
\end{equation}%
as has been demonstrated by Haroche and co-workers \cite{haroche}. Let the
distance between the two mirrors M1 and M2 be $L$, which, for the lowest
mode of the resonator, will be around $\lambda /2$, i.e., half a microwave
wavelength. We shall apply the method of images to solve this problem.

The magnitude of the instantaneous velocity of mirror M2 moving to the right
relative to mirror M1 will be given by%
\begin{equation}
v=\dot{L}
\end{equation}%
Using the mathematical method of induction in conjunction with the physical
method of images, one finds that the positions $z^{(2n+1)}$ of successively
higher odd order images M1$^{(2n+1)}$, $(n=0,1,2...)$, of the mirror M1 on
the right side of mirror M2, as seen by the eye in Figure 1, will be given by%
\begin{equation}
z^{(2n+1)}=(n+2)L\text{ , }(n=0,1,2...)
\end{equation}%
Therefore the magnitudes of the velocities of the successively higher-order
images of mirror M1 on the right side of mirror M2 will be given by%
\begin{equation}
v^{(2n+1)}=\dot{z}^{(2n+1)}=(n+2)\dot{L}=(n+2)v\text{ , \ }(n=0,1,2...)
\end{equation}%
One concludes that the velocity of the higher-order images to the right of
M2 will steadily increase as the order $n$ of the images M1$^{(2n+1)}$
steadily increases.

Now the meaning of the quality factor $Q$ of the resonator is that it is roughly the number
of back-and-forth reflections of a ray between mirrors M1 and M2 before the
ray escapes the resonator. Hence let us set $n\sim Q$ for the total number
of bounces of a ray within the Fabry-Perot configuration of Figure 1, before
it leaves the cavity. A rough estimate of the magnitude of the velocity of
the highest order image of M1 just before the ray escapes from the
resonator, will therefore be on the order of%
\begin{equation}
v^{(2Q+1)}=\dot{z}^{(2Q+1)}=(Q+2)\dot{L}=(Q+2)v\sim Qv
\end{equation}%
where $Q\sim 10^{10}$. Therefore, although the mirror M2 may initially be
moving nonrelativistically with $v<<c$, nonetheless after $Q$ reflections
within the Fabry-Perot, the effective velocity of the $Q$th order image of
M1, namely M1$^{(2Q+1)}$, can become \emph{effectively} relativistic.

For example, if mirror M2 initially moves with a nonrelativistic velocity $%
v\sim 1$ cm s$^{-1}$, then after $Q\sim 10^{10}$ reflections, the $Q$th
order image of M1 will appear to be moving with a relativistic velocity on
the order of $v^{(2Q+1)}\sim 10^{10}$ cm s$^{-1}$. Since the
relativistically moving images of M1 will be carrying both a charge and a
mass with them as they move, these images can radiate significant amounts of
both electromagnetic and gravitational radiation \cite{thomson}.

One concludes that it is not necessary for the center of mass of quantum
mesoscopic objects to move near the speed of light in order for them to emit
appreciable amounts of radiation. Rather, it is sufficient for their \emph{%
images} to \emph{effectively} move relativistically inside an extremely high 
$Q$ resonator, such as the SC resonator in Figure 1. 

Let us estimate the order of magnitude of the gravitational wave power
emitted in the quadrupolar radiation arising from the moving masses of the
images of Figure 1. We shall assume at the outset that the SC
\textquotedblleft mirror\textquotedblright\ boundary conditions of \cite%
{physica e}\ are valid here, which allows us then to use the method of
images. But first let us estimate the amount of electromagnetic wave power
emitted by the moving charges of the images of Figure 1, starting from the
following formula for the emission of quadrupole radiation given by \cite
{Jackson}
 
\begin{equation}
\mathcal{P}^{\text{(quad)}}=\frac{Z_{0}\omega ^{6}}{1440\pi c^{4}}%
\sum_{ij}\left\vert \mathcal{Q}_{ij}\left( \omega \right) \right\vert ^{2}
\label{Jackson's quadrupole formula}
\end{equation}
where $Z_{0}=377$ ohms is the characteristic impedance of free space, and
where $\mathcal{Q}_{ij}\left( \omega \right) $ is a component of the
electric quadrupolar moment tensor of a radiating source which is
oscillating at an angular frequency $\omega $. We shall apply (\ref%
{Jackson's quadrupole formula}) to the radiation emitted by the dominant
oscillating quadupole moment as seen by the eye in Figure 1, which will
arise from the moving image charge M1$^{\text{(2}n\text{+1)}}$, relative to
the moving image M1$^{\text{(1)}}$, when $n\approx Q$, the quality factor of
the resonator. Now the dominant component of the electrostatic quadrupole
moment $\mathcal{Q}_{zz}^{\text{(stat)}}$ as seen by the eye in Figure 1
arising from the charge $+q$ on the mirror image M1$^{\text{(2}Q\text{+1)}}$%
, relative to the image charge $+q$ on the mirror image M1$^{\text{(1)}}$,
will be given by%
\begin{equation}
\mathcal{Q}_{zz}^{\text{(stat)}}\approx qd_{0}^{2}
\end{equation}%
where $d_{0}\approx QL\approx Q\lambda /2$ is the distance from M1$^{\text{(2%
}Q\text{+1)}}$ to M1$^{\text{(1)}}$, and where $L=\lambda /2$ is a microwave
half-wavelength. Let the motion of the mirror M2 be given by a small,
sinusoidally varying displacement $\Delta d\left( t\right) =\varepsilon
_{\max }\sin \omega t$ with a small displacement amplitude $\varepsilon
_{\max }$. It follows that the time-varying component of the dominant
quadrupole moment formed by the images M1$^{\text{(1)}}$ and M1$^{\text{(2}Q%
\text{+1)}}$ will be%
\begin{equation}
\Delta \mathcal{Q}_{zz}\left( t\right) \approx 2qQd_{0}\varepsilon _{\max
}\sin \omega t
\end{equation}%
From (\ref{Jackson's quadrupole formula}), one then obtains an estimate of
the time-averaged emitted power%
\begin{equation}
\mathcal{P}^{\text{(quad)}}\approx \frac{\pi }{720}\frac{Z_{0}\omega ^{4}}{%
c^{2}}q^{2}Q^{4}\varepsilon _{\max }^{2}
\label{Quad power in terms of epsilon}
\end{equation}%
as seen by the eye in Figure 1. Now the on-resonance solution to the simple
harmonic oscillation equation of motion of the moving mirror M2 with a mass $%
m$ is given by%
\begin{equation}
\varepsilon _{\max }=\frac{qE_{\max }}{m\omega ^{2}}Q
\label{epsilon related to max electric field}
\end{equation}%
where $E_{\max }$ is the maximum amplitude of the longitudinal electric
field being applied to the electrostatic charge $+q$ of M2 in the transverse
magnetic mode. From (\ref{Quad power in terms of epsilon}) and (\ref{epsilon
related to max electric field}), we see that the emitted power in quadrupole
radiation from the images M1$^{\text{(1)}}$ and M1$^{\text{(2}Q\text{+1)}}$
will be proportional to the power injected into the resonator which is
driving the motion of the mirror M2. Let $\mathcal{P}_{i}$ be this injected
power. This leads to the relationship%
\begin{equation}
E_{\max }^{2}=\frac{Z_{0}Q}{\mathcal{A}_{\text{eff}}\pi }\mathcal{P}_{i}
\end{equation}%
where $\mathcal{A}_{\text{eff}}\approx \lambda ^{2}$ is the effective
cross-sectional area of the transverse magnetic mode of the resonator. One
then finds that%
\begin{equation}
\mathcal{P}^{\text{(quad)}}=\eta _{\text{eff}}\mathcal{P}_{i}\text{ \ with \ 
}\eta _{\text{eff}}\approx \frac{Z_{0}^{2}q^{4}Q^{7}}{720m^{2}c^{2}\lambda
^{2}}
\end{equation}%
where $\eta _{\text{eff}}$ is the effective conversion efficiency from $%
\mathcal{P}_{i}$ into $\mathcal{P}^{\text{(quad)}}$. Now the maximum
possible value of the effective conversion efficiency $\eta _{\text{eff}}$
will be unity. Therefore the electrostatic charge $q$ deposited onto M2
needed to achieve $\eta _{\text{eff}}\approx 1$ is therefore on the order of%
\begin{equation}
q\left( \eta _{\text{eff}}\approx 1\right) \approx \left( \frac{%
720m^{2}c^{2}\lambda ^{2}}{Z_{0}^{2}Q^{7}}\right) ^{1/4}
\end{equation}%
which, when evaluated for the following parameters,%
\begin{eqnarray}
m &=&m_{\text{Planck}}=21.8\text{ micrograms} \\
\lambda  &=&1.5\text{ centimeters (i.e., 20 GHz)} \\
Q &\approx &10^{9}
\end{eqnarray}%
yields an order-of-magnitude value for the required electrostatic charge to
be deposited onto M2 which is given by%
\begin{equation}
q\left( \eta _{\text{eff}}\approx 1\right) \approx 1.4\times 10^{-17}\text{
Coulombs}\approx 93\text{ electrons}
\end{equation}%
which should be easily feasible. The resulting charge-to-mass ratio of the
moving mirror M2 is%
\begin{equation}
\frac{q\left( \eta _{\text{eff}}\approx 1\right) }{m}\approx 6\times 10^{-10}%
\text{ Coulombs kg}^{-1}
\end{equation}%
which is within an order of magnitude of the \textquotedblleft
criticality\textquotedblright\ charge-to-mass ratio (\ref{criticality
charge-to-mass ratio}) (see \cite{charge-to-mass criticality}), which is the
condition that\ when two identical charges $q$ are attached to two identical
masses $m$ moving anti-symmetrically with respect to each other in
sinusoidal motion,\thinspace\ this pair of particles will radiate equal
amounts of electromagnetic and gravitational wave power. This indicates that the amount of
quadrupolar gravitational radiation emitted by the configuration of image
masses in Figure 1 can easily be made comparable to the amount of
quadrupolar electromagnetic radiation emitted by the image masses of this
configuration by adjusting the amount of electrostatic charge deposited on
M2. Therefore the configuration sketched in Figure 1 could become the basis
for a design of an efficient converter (i.e., transducer) from electromagnetic to gravitational
radiation, and vice versa.


\begin{thebibliography}{9}
\bibitem{FQMT11} R.Y. Chiao, L.A. Martinez, S.J. Minter, A. Trubarov,
\textquotedblleft Parametric oscillation of a moving mirror driven by
radiation pressure in a superconducting Fabry--Perot resonator
system,\textquotedblright\ Phys. Scr. T \textbf{151}, 014073 (2012); arXiv:
1207.6885.

\bibitem{Aharonov MS} R.Y. Chiao, R.W. Haun, N.A. Inan, B.S. Kang, L.A.
Martinez, S.J. Minter, G.A. Mu\~{n}oz, and D.A. Singleton, \textquotedblleft 
{A gravitational Aharonov-Bohm effect, and its connection to parametric
oscillators and gravitational radiation,}\textquotedblright\ (Aharonov 80th
birthday Festschrift article), arXiv: quant-ph 1301.4297. When the charge $q$
on the mirrors is sufficiently large, and when a transverse magnetic mode of
the resonator is being excited, then the motion of a thin, flexible mirror
will be \textquotedblleft slaved\textquotedblright\ to that of the
longitudinal electric field of this transverse magnetic mode, so that the
mirror will be forced to move at the microwave resonance frequency of this
mode instead of its acoustical resonance frequency.

\bibitem{haroche} S. Kuhr, S. Gleyzes, C. Guerlin, J. Bernu, U.B. Hoff, S.
Del\'{e}glise, S. Osnaghi, M. Brune, J.M. Raimond, S. Haroche, E. Jacques,
P. Bosland, and B. Visentin, \textquotedblleft Ultrahigh finesse Fabry-Perot
superconducting resonator,\textquotedblright\ Appl. Phys. Lett. \textbf{90},
164101 (2007).

\begin{figure}[ptb]
\begin{center}
\includegraphics[width=5in]{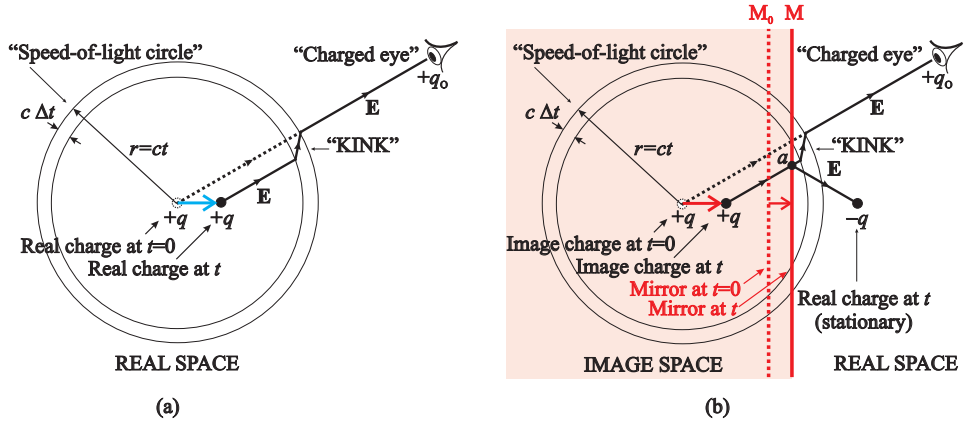}
\end{center}
\caption{(a) Snapshot of a charged eye looking at a \textquotedblleft kicked\textquotedblright\ charge. A \textquotedblleft kink\textquotedblright\ develops in an electric field line $\mathbf{E}$ emanating from the charge at the \textquotedblleft speed-of-light circle,\textquotedblright\ which represents the radiation from the kicked charge. (b) Snapshot of a charged eye looking at a \textquotedblleft kicked\textquotedblright\ mirror. Again, a
\textquotedblleft kink\textquotedblright\ develops at the \textquotedblleft
speed-of-light circle\textquotedblright\ representing radiation from the
\textquotedblleft kicked\textquotedblright\ mirror.}
\end{figure}

\bibitem{thomson} Based on Thomson's model for radiation from an accelerated
charge due to a propagating \textquotedblleft kink\textquotedblright\
superimposed upon a line of the Coulomb field of the charge that forms at
the \textquotedblleft speed-of-light circle\textquotedblright\ (see Figure
2(a)), one infers that an accelerated image charge (see Figure 2(b)),
which is seen by a stationary observer as the image of a stationary charge $%
-q$ observed within the accelerated mirror, can emit radiation just as
surely as an accelerated charge would radiate as it moves towards a
stationary mirror. It is the conducting boundary conditions of the
\textquotedblleft kicked\textquotedblright\ mirror at point $a$ on the
surface of the moving mirror M in Figure 2(b) that leads to the formation of
Thomson's \textquotedblleft kink\textquotedblright\ by the charge $+q$ of
the accelerated image as seen within the accelerated mirror by the observer.
Note that the original, real charge $-q$ is stationary and not moving in
Figure 2 (b). Nevertheless, radiation will be produced by the time-varying
surface currents induced in the \textquotedblleft kicked\textquotedblright\
mirror by the stationary charge $-q$. By the relativity of motion, it cannot
matter whether it is the mirror that is moving towards a stationary charge,
or that it is the charge that is moving towards a stationary mirror.
Therefore the energy produced in the radiation associated with the
\textquotedblleft kink\textquotedblright\ by the moving mirror in the
presence of the stationary charge, must be exactly the same as the energy
produced in radiation associated with the \textquotedblleft
kink\textquotedblright\ by the moving charge in the presence of the
stationary mirror. Since we know that energy will be radiated away in the
latter case, it follows that energy must also be radiated away in the former
case. Therefore accelerated image charges must radiate. Similarly, the
superconducting \textquotedblleft mirror\textquotedblright\ boundary
conditions found in \cite{physica e} will lead to the formation of Thomson's
\textquotedblleft kink\textquotedblright\ by the mass\emph{\ }of a moving
image as seen within a moving mirror by a stationary observer. Therefore
energy will be radiated away in the form of gravitational radiation by an accelerated
SC mirror moving relative to a stationary mass, just as surely as an
accelerated mass moving relative to a stationary mirror would radiate away
energy in this form of radiation.

\bibitem{physica e} S.J. Minter, K. Wegter-McNelly, and R.Y. Chiao,
\textquotedblleft Do mirrors for gravitational waves
exist?\textquotedblright , Physica E \textbf{42}, 234 (2010);
arXiv:0903.0661.

\bibitem{Jackson} J.D. Jackson, \emph{Classical Electrodynamics}, 4th edition, (John Wiley \& Sons,1999), p.415, Eq (9.49).

\bibitem{charge-to-mass criticality} If the charge-to-mass ratio of two
identical charges $q$ attached to two identical masses $m$, which are moving
anti-symmetrically with respect to each other in sinusoidal motion, were to
be adjusted to the \textquotedblleft criticality\textquotedblright\ value 
\cite{Chiao-Townes-volume}%
\begin{equation}
\left( \frac{q}{m}\right) _{\text{criticality}}=\sqrt{4\pi \varepsilon _{0}G}%
=8.6\times 10^{-11}\text{ C/kg}  \label{criticality charge-to-mass ratio}
\end{equation}%
then equal amounts of electromagnetic and gravitational radiation would be produced by the
sinusoidal motion of this pair of particles (see text above). Thus an
efficient transducer between gravitational and electromagnetic radiation could be constructed.

\bibitem{Chiao-Townes-volume} R.Y. Chiao, \textquotedblleft New directions
for gravitational-wave physics via `Millikan oil drops',\textquotedblright\
in the Festschrift for Charles H. Townes, \textit{Visions of Discovery},
edited by R.Y. Chiao, M.L. Cohen, A.J. Leggett, W.D. Phillips, and C.L.
Harper, Jr. (Cambridge University Press, 2011), p. 348,
arXiv:gr-qc/0904.3956.
\end{thebibliography}
\end{document}